\documentclass[runningheads,a4paper]{llncs}
\usepackage{amsmath}
\usepackage{amssymb}
\usepackage{graphicx}

\usepackage{url}
\urldef{\mailsa}\path|raffaele.marino@unifi.it|

\begin{document}

\mainmatter  

\title{Where do hard problems really exist?}

\titlerunning{Where do hard problems really exist?}

%
%
\author{Raffaele Marino}
\authorrunning{Raffaele Marino}
\institute{Dipartimento di Fisica e Astronomia,\\
Via Giovanni Sansone 1, Sesto Fiorentino, Firenze, Italy\\
\mailsa\\}

%
%

\maketitle

\begin{abstract}
This chapter delves into the realm of computational complexity, exploring the world of challenging combinatorial problems and their ties with statistical physics. Our exploration starts by delving deep into the foundations of combinatorial challenges, emphasizing their nature. We will traverse the class P, which comprises problems solvable in polynomial time using deterministic algorithms, contrasting it with the class NP, where finding efficient solutions remains an enigmatic endeavor, understanding the intricacies of algorithmic transitions and thresholds demarcating the boundary between tractable and intractable problems. We will discuss the implications of the P versus NP problem, representing one of the profoundest unsolved enigmas of computer science and mathematics, bearing a tantalizing reward for its resolution.
Drawing parallels between combinatorics and statistical physics, we will uncover intriguing interconnections that shed light on the nature of challenging problems. Statistical physics~
unveils profound analogies with complexities witnessed in combinatorial landscapes.
Throughout this chapter, we will discuss the interplay between computational complexity theory and  statistical physics. By unveiling the mysteries surrounding challenging problems, we aim to deepen understanding of the very essence of computation and its boundaries.
Through this interdisciplinary approach, we aspire to illuminate the intricate tapestry of complexity underpinning the mathematical and physical facets of hard problems.

\end{abstract}

\tableofcontents
\clearpage
\section{Introduction}
\label{sec::intro}
Life is a splendid ballet of decisions and consequences, an intricate tapestry woven from the threads of choices. Our desires, as diverse and dynamic as they are, often find themselves in a bitter symphony of conflict. Consider a situation in your life where you desire to maintain a close friendship with two individuals, say, Ms. Red and Ms. Blue \cite{mezard1987spin}. However, these two harbor a bitter animosity towards each other - a predicament that could be lifted straight from a classical tragedy. The task of preserving friendships with both becomes a delicate dance, a strenuous tightrope walk. This already perplexing scenario quickly turns into a formidable puzzle as the number of individuals and their relationships with each other expand.

Let's transpose this situation into a less dramatic, yet equally challenging context: a social gathering, a party filled with an eclectic mix of people. You are one among the $N$ guests, and you have been tasked with a unique mission - find the largest clique within the party, the biggest group of individuals who all know each other personally. Imagine drawing a map of the party, with each guest represented by a point and a line connecting those who know each other. This clique problem is akin to finding the most tightly-knit friend group in the party. It sounds simple, but even for a modest number of guests, the possibilities to explore multiply rapidly, and the problem becomes computationally challenging  \cite{karp2010reducibility}. 

Let's extend our party scenario a step further. Imagine the host, keen on ensuring everyone enjoys the event, is aware of certain preferences among the guests. Some guests, for instance, would rather not be at the same table with certain others, while some wish to be seated near the band, and others away from it. The host's challenge is to accommodate everyone's preferences as much as possible. This is a classic example of what is known in computer science as a Satisfiability (SAT) problem \cite{gu1996algorithms}. The SAT problem is a decision problem in the realm of propositional logic. Each guest's preference can be represented as a clause of  literals (specific conditions to be satisfied). The problem is to determine if there is an assignment of truth values (in our case, seating arrangements) that satisfies all the clauses (all guest preferences). Although our party scenario simplifies the complexities of the SAT problem, it captures its essence: finding a solution within a sea of possibilities, under specific constraints. Whether we are assigning truth values to variables in propositional logic or navigating the intricate preferences of party guests, we are exploring the captivating world of combinatorial problems.

As we progress through our exploration, remember our party, its social network, its cliques, and seating arrangements. These real-life scenarios, imbued with the complexities of combinatorial problems, form a bridge between abstract computational theory and the intricate dynamics of the world around us, revealing an intimate interplay between information \cite{shannon1993claude,richardson2008modern} and statistical physics \cite{ma1985statistical,huang2009introduction}, illuminating the astounding complexity \cite{GiorgioParisi1987,cugliandolo2023scientific,parisi2023thoughts} of the world around and within us.

\section{Some basic definitions of Graph Theory}
\label{sec::graph_theory}
As stated in the introduction, we can draw a map of the party, with each guest represented by a point, and a line connecting those who know each other. In this case, what we are doing is making an abstraction of the problem. By considering each guest as a point and each friendly interaction as a line, we are defining a \textbf{graph} \cite{bollobas1998modern}.

A graph $\mathcal{G}$, in the mathematical sense, is a set of objects referred to as vertices (or nodes), and a set of edges (or arcs), formally defined as $\mathcal{G}=(\mathcal{V},\mathcal{E})$ where $\mathcal{V}$ is a non-empty set of vertices and $\mathcal{E}$ is a set of unordered pairs of distinct vertices $\mathcal{E} \subseteq \mathcal{V}\times \mathcal{V}$. For instance, an edge joining vertices $i$ and $j$ is identified as $e=(i,j)$.  The vertex set $\mathcal{V}$ can be any  finite set but one often takes the set of the first $|\mathcal{V}|$ integers $\mathcal{V}=\{1,2,...|\mathcal{V}|\}$. In our party scenario, each guest becomes a vertex, and each friendly connection morphs into an edge. The complete web of connections between all guests forms the graph of the party. A \textbf{weighted graph} is a graph where a cost (a real number) is associated with every edge. An \textbf{undirected graph} is a graph in which edges have no direction. The edges imply a two-way relationship, in that each edge can be traversed in both directions. A \textbf{directed graph}, also called a digraph, is a graph in which edges have orientations. The edges represent a one-way relationship, in that each edge can be traversed in a single direction from one vertex to another. A graph can be effectively represented using a mathematical structure called an adjacency matrix. An adjacency matrix is a square matrix where each entry indicates whether pairs of vertices are adjacent or not in the graph. In the context of an unweighted, undirected graph, the adjacency matrix is binary. That is, if there is a link between node $i$ and node $j$, the entry $(i, j)$ in the matrix will be $1$, otherwise, it will be $0$.  If the graph is weighted, the weights can be used to fill the matrix rather than mere binary values. The adjacency matrix provides a powerful tool for graph analysis, with the matrix operations enabling efficient computation of various graph properties. 

The \textbf{degree} of a vertex is the number of edges connected to it. A \textbf{path} in a graph would represent a sequence of vertices, such that from each of its vertices there is an edge to the next vertex in the sequence. Formally, a path of length $n$ in a graph is a sequence $P = v_0, e_1, v_1, e_2, v_2, ..., e_n, v_n$ of vertices and edges such that for $1 \leq i \leq n, e_i$ is the edge connecting $v_{i-1}$ and $v_i$. A graph is \textbf{connected} if for every pair of vertices, there is a path which connects them. A complete graph on $N$ vertices, denoted by $K_N$, is a simple undirected graph in which every pair of distinct vertices is connected by a unique edge. In this graph, the degree of every vertex (the number of edges connected to it) is $N-1$.

A \textbf{clique} in a graph $\mathcal{G}$ is a subset $K_n$ of $n$  vertices of $\mathcal{G}$ such that every two distinct vertices in $K_n$ are adjacent (connected by a common edge). It is our tight-knit group of friends, where each member knows every other member. A \textbf{cycle} $\mathcal{R}$ in graph theory is a path that starts and ends at the same vertex, without revisiting any other vertices along the way. Formally, a cycle of length $n$ in a graph is a sequence $\mathcal{R} = v_0, e_1, v_1, e_2, v_2, ..., e_n, v_0$ of vertices and edges such that $P = v_0, e_1, v_1, e_2, v_2, ..., e_n, v_n$ is a path, the edges $e_1, e_2, ..., e_n$ are all distinct, and the vertices $v_0, v_1, ..., v_{n-1}$ are all distinct. A \textbf{tree}, on the other hand, is a special kind of graph that has no cycles and is connected. This means in a tree on $N$ vertices, there are exactly $N-1$ edges, and there exists a unique path between every pair of vertices.

By recasting the social dynamics of our party into the language of graph theory, we gain powerful tools to explore and analyze the complexities of the party's social fabric. Yet, the underlying principles remain relatable and intuitive, grounded in the everyday experiences of social interaction.

\section{Combinatorial Optimization Problems}
\label{sec::COP}
At its core, a combinatorial problem involves searching for an optimal solution from within a finite, but typically very large, set of possible solutions. It is, essentially, an exploration of the landscape of possibilities. These problems arise in a myriad of contexts, from scheduling airline flights to arranging chess pieces \cite{papadimitriou1998combinatorial}.


In more formal terms, combinatorial problems involve finding ways to combine elements in a set satisfying certain rules. These rules might require us to arrange the elements in a sequence, select a group without considering the order (like choosing a committee from a larger group), or create more complex structures like graphs or networks \cite{schneider2007stochastic}. A subset of combinatorial problems, known as combinatorial optimization problems, involve finding the \textit{best} solution from the set of all possible solutions \cite{korte2011combinatorial}. What defines \textit{best} can vary. In a route planning problem, for instance, \textit{best} might mean the shortest route; in a resource allocation problem, \textit{best} might mean maximizing profit or minimizing cost. 
Combinatorial optimization problems are, therefore, a class of computational problems that involve finding the best solution among a finite set of possible solutions. These problems often arise in various fields, such as operations research, logistics, scheduling, and network design. They typically involve making choices or selecting elements from a discrete set in order to optimize an objective function.

In a combinatorial optimization problem an \textbf{instance} is described by a finite set $\chi$ of allowed configurations, and an \textbf{objective function} $E$ defined on this set and taking values in $\mathbb{R}$. The optimization problem consists in finding the \textbf{optimal configuration} $C\in \chi$, namely the configuration that minimizes the objective function $E(C)$. Any set of such instances defines a combinatorial optimization problem.

For solving an optimization problem we need an algorithm. For explaining what an algorithm is, it is useful considering again the challenge at our party described in the introduction: finding the largest group of friends who all know each other - the maximum clique.  To tackle this problem, we could approach it like a strategic party planner, devising a methodical process to identify the best possible group.

An algorithm, in this context, can be thought of as that party planner's strategy. It's a structured set of instructions that sifts through the web of friendships at the party, comparing, evaluating, and remembering which groups of friends satisfy our criteria of all knowing each other. The algorithm does not just wander aimlessly through the party guests, it navigates with purpose. It might start with one individual, list down all their acquaintances, and then repeat the process for each acquaintance, systematically building up groups. It keeps track of the largest group it has found so far, and with each new group it forms, it checks - is this group larger? If so, it remembers this new group as the current best. As the evening progresses, the algorithm continues its work, tirelessly and methodically exploring the social network. The \textit{best} group it identifies might change multiple times throughout the process, but by the time the party ends, the algorithm has done its job: it has found the largest group of mutual friends. In essence, an algorithm is our strategic tool in the quest for optimal solutions. It guides us through the vast solution space with a plan, a set of rules, and the aim of optimizing a specific measure - in our case, the size of the friend group.

In more general words,  an algorithm solves an optimization problem if, for every instance of the optimization problem, it gives the optimal configuration or it computes the minimum of the objective function $E$. In all these problems there is a natural measure of the size of the problem, which is $N$, thus the number of variables used to define a configuration of the problem. In the context of the challenge at our party the size of the problem is $N$, the number of guests.

Combinatorial optimization problems come in many forms, each with its own set of constraints and objectives. Here are a few well-known examples:
\begin{itemize}
    \item \textbf{Traveling Salesman Problem (TSP)} \cite{flood1956traveling}: This hypothetical problem paints a picture of a traveling salesperson who must navigate through a number of different cities, visiting each city only once and then returning home, all while keeping the total distance traveled as short as possible.  Given a list of cities and the distances between each pair of cities, the objective is to find the shortest possible route that visits each city exactly once and returns to the starting city. While it may seem simple, the TSP is a notoriously complex problem. As the number of cities increases, the number of possible routes grows exponentially, making it virtually impossible to solve by simply checking all possible routes.
    \item \textbf{Graph Coloring Problem} \cite{jensen2011graph}: The problem objective is to assign a color to each vertex such that no two adjacent vertices share the same color, while using the minimum number of colors, given an undirected graph. This might remind you of a coloring book, where you want to color a map so that no two neighboring regions have the same color. It seems simple, but as the size and complexity of the graph increase, finding the minimum number of colors that fulfills this condition - often referred to as the graph's chromatic number - can be a daunting task.
    \item \textbf{Max-Cut} \cite{alon_krivelevich_sudakov_2005}: The Max-Cut problem can be stated as follows: "Given a graph, how can you divide or \textit{cut} the vertices into two separate groups so that there are as many edges as possible connecting the two groups?". Imagine a group of friends who want to split into two teams for a game. They want to arrange the teams so that as many friends as possible have a friend on the other team. The Max-Cut problem is like this scenario: each friend is a vertex, each friendship is an edge, and the problem is finding the best way to form the teams (the \textit{cut}) to maximize friendships between the teams.
    \item \textbf{Maximum Clique Problem} \cite{bomze1999maximum}: As stated in the introduction, the Maximum Clique Problem  can be framed as follows: "Given a graph, can you find the largest group of vertices where each vertex is connected to every other vertex in the group?". 
    Really close to Maximum Clique Problem is the Maximum Independent Set Problem.  This problem is about finding the biggest group of points in a graph, where none of the points in the group are directly connected to each other. 
    \item \textbf{SAT Problem} \cite{biere2009handbook}: The SAT problem, short for Satisfiability, is a classic question in the world of computer science, specifically in the realm of combinatorial optimization and computational complexity. The problem can be presented like this: "Given a Boolean expression consisting of variables that can be either true or false, combined using AND, OR, and NOT operators, can you assign values to the variables in a way that makes the entire expression true?" This is the general SAT problem. 
\end{itemize}

Only a few examples of combinatorial optimization problems are given above, but there are  many more \cite{garey1979computers}. In this list, however we have not yet described how and where hard problems are. We need before understanding how complex these problems are, and therefore we need a definition of the measure of the \textbf{complexity}.

\section{Computational Complexity and Complexity Classes}
\label{sec::Comp_comp_p_vs_NP}

The exploration of complexity, or the inherent intricacy of interconnected systems, manifests not only in contemporary fields such as computational sciences and mathematics but also dates back to the philosophical discourses of Aristotle and Thomas Aquinas. As we delve into the realms of computational complexity and complexity classes, it is worth recognizing these philosophical underpinnings to understand the true nature of complexity.

The philosophical question of the "\textit{one and the many}" provides an apt metaphor for computational complexity. As Aristotle contemplates the emergence of a unified whole from its individual components, we also consider how simple computational elements contribute to complex tasks. Similar to Aquinas' view of society as an ontological entity, computational systems could be conceived as a unified entity - a complex that is more than the sum of its individual computational parts \cite{fazio2009}. This entity demonstrates emergent properties, which allow for the achievement of computational goals, unattainable by individual computational components. Taking a cue from the physical sciences, let's consider the example of a heavy hydrogen molecule (D$_2$). It consists of the same basic particles as a helium atom, but the organization and interactions of these particles produce entirely different properties. Similarly, the realm of computational systems and their complexity does not merely lie in the count of computational parts. It lies in the configuration, coordination, and communication of these parts, which inevitably leads to emergent properties, behavior, and tasks of various complexity.

The nuance between \textit{complex} and \textit{complicated} has a profound implication in understanding computational complexity. While both terms suggest a certain level of intricacy, \textit{complex} denotes interconnectivity to form a whole, while \textit{complicated} indicates multiple layers folded onto themselves. This distinction is not merely semantics; it holds the key to our understanding of computational complexity. A computational problem may be complicated, having several layers of calculations, but its complexity lies in how these calculations are interwoven, and how they form a unified task.


As we tread further into the concepts of computational complexity and complexity classes, we will unfold the idea of how simple, individual computational tasks, much like the single members of Aquinas' society or Aristotle's parts of a whole, interweave into a complex network of processes. This network, while potentially complicated, ultimately reveals the essence of computational complexity – the unity achieved by the system.

Embracing the common consensus, we gauge the complexity of an algorithm by enumerating the "\textit{elementary operations}" requisite for problem-solving. Typically, we place our emphasis on the asymptotic behavior of this quantity as $N$ approaches infinity. Undeniably, designing algorithms with the least possible complexity holds immense practical significance. This approach is frequently employed in the resolution of combinatorial optimization problems.

Generally, we categorize problems into three fundamental types: Firstly, the \textbf{optimization} problem, where the objective is to pinpoint an optimal configuration, denoted as $C^*$. Secondly, the \textbf{evaluation} problem, which involves calculating the cost $E(C^*)$ associated with this optimal configuration. Lastly, we have the \textbf{decision} problem, where the query to be addressed is, "Does a configuration exist with a cost less than a specified value $E_0$?"

\subsection{Some basic elements on the theory of computational complexity}
\label{subsec::compl_theory}
The exploration of computational complexity \cite{mezard2009information} is a crucial domain within the theoretical aspects of computer science, aspiring to lay the groundwork for an intrinsic understanding of the subject. An essential part of this exploration entails establishing a relative scale of complexity - an understanding of which problems are more challenging, or \textit{harder}, than others. In this context, a \textit{hard problem} is typically interpreted as one necessitating a longer computational time for its resolution.

To address the notion of computational complexity with rigor, one might feel inclined to define a precise model of computation, as initially proposed by Alan Turing \cite{turing2009computing}.  He showed that some questions of the possibility of computation could be made concrete, and some even answered, but introducing a concrete model of a universal computer, which has ever since been called a Turing machine.  Many such models of a basic computer now exist, some of them real and not just mathematical ideals, but Turing's model is just about the simplest and provides a reference. However, establishing such a model may lead us down a considerably intricate path, one that may be too vast to explore in this context. Rather, we adopt a more informal yet fundamentally sound approach for our current purpose, where we assess the running time of an algorithm based on \textit{elementary operations}. Elementary operations are the fundamental operations executed by an algorithm, including, but not limited to, comparisons, sums, multiplications, and so forth. By observing how these basic operations scale with the input size, we can grasp the essence of an algorithm's complexity. A critical assumption in this context is that the size of the operands remains uniformly bounded. In other words, regardless of the complexity or scale of the problem at hand, we assume that each elementary operation can be performed within a fixed amount of time.

This approach, although informal, offers a practical, intuitive, and broadly correct method to assess and compare computational complexity. With it, we can begin to understand the hierarchy of problems, ranging from simpler ones that require fewer elementary operations, to harder problems that demand significantly more computational resources. Keep in mind that this understanding is a cornerstone in the theory of computational complexity. It allows us to discern the intricate landscape of problems and algorithms, setting a framework where we can make meaningful comparisons and draw insightful conclusions about their relative complexities. By building upon this foundation, we can then venture deeper into the theory, unraveling more sophisticated concepts, and further sharpening our understanding of this crucial aspect of computer science. Building upon the foundational understanding of computational complexity, we are led to the concept of \textit{worst-case scenario} \cite{arora2009computational}. This concept becomes pivotal when we consider that for a given algorithm, particularly one designed to solve a combinatorial optimization problem, its running time can significantly vary across different instances, even if these instances are of the same size. A natural question then arises: how can we quantify the overall difficulty or \textit{hardness} of such a problem?
By focusing on this case, we set a benchmark of hardness for the problem, ensuring that our algorithm can handle the most challenging instances. This approach, although perhaps seeming pessimistic, offers a form of guarantee on the algorithm's performance, providing an upper limit on the running time. Let's consider a simple example to illustrate this: the problem of sorting a list of numbers \cite{cormen2022introduction}. Given an algorithm that sorts this list, the running time can vary depending on the initial order of numbers. If the list is already sorted (the best-case scenario), the algorithm might finish its job very quickly. However, if the list is sorted in reverse order (often the worst-case scenario for many sorting algorithms), it would take considerably longer for the algorithm to sort it. When quantifying the hardness of the sorting problem, we consider the running time of the algorithm in this worst-case scenario, the reversed list. This approach gives us a robust measure of the problem's complexity, preparing us for the most demanding instances we might encounter. Hence, the worst-case scenario analysis serves as a protective buffer, ensuring that our algorithm is equipped to handle even the most difficult cases, thereby guaranteeing its effectiveness across a wide range of problem instances.

Adopting this worst-case scenario strategy garners two significant benefits: Firstly, it facilitates the construction of a \textit{universal} theory. By focusing on the hardest instances, we ensure that our understanding and analysis apply across the entire spectrum of problem instances, thereby fostering a comprehensive and universal theory of computational complexity. This universality allows us to compare different algorithms and problems on a common ground, regardless of their specificities. Secondly, once we have gauged the worst-case running time for a given algorithm, we establish a performance guarantee that applies to any instance of the problem. This guarantee serves as an assurance that no matter the specific details of a problem instance, the time taken by the algorithm will not exceed the worst-case estimate. This can be invaluable in real-world applications where predicting the exact nature of problem instances might not be feasible, but having an upper bound on the algorithm's running time can aid in effective planning and resource allocation. Thus, the emphasis on the worst-case scenario not only aids us in crafting a comprehensive theory of computational complexity but also offers practical benefits by providing performance guarantees, enhancing the reliability and predictability of our computational solutions.

Navigating through the fascinating terrain of computational complexity, we encounter two critical signposts that guide us in categorizing algorithms: Polynomial and Super-polynomial time complexities. These distinct categories function as valuable scales, assisting us in distinguishing between various problem types based on how the computational requirements inflate with an increase in the size of the instances \cite{mezard2009information}.
\begin{itemize}
    \item \textbf{Polynomial Time Complexity}: An algorithm is said to be Polynomial if the computational time it demands, represented as $T(N)$, can be bounded from above by a fixed polynomial in the size $N$ of the instance. Mathematically, if there exists a non-negative integer $k$ such that $T(N) = O(N^k)$ for all $N$ large enough, we say that the problem exhibits polynomial time complexity.
    \item  \textbf{Super-polynomial Time Complexity}: Conversely, an algorithm enters the Super-polynomial time complexity category if its running time, $T(N)$, can't be upper-bounded by any fixed polynomial in the size of the instance. This is, for instance, the case if the time grows exponentially with the size of the instance (it is used to call these algorithms \textbf{exponential}).
\end{itemize}

The concept of the \textit{size} of a problem, however, remains somewhat nebulous in our discussion thus so far. One might naturally question whether a shift in the definition of \textit{size} could potentially reclassify a problem from polynomial to super-polynomial, or vice versa. To illustrate this, let's examine the assignment problem \cite{caracciolo2021criticality,caracciolo2017finite} involving $2N$ points. The size of this problem could be defined as $N$, $2N$, or even $N^2$. Yet, all these definitions of size can be seen as polynomial functions of one another. Hence, the choice among these definitions does not alter the categorization of an algorithm as either polynomial or super-polynomial. These are considered \textit{natural} definitions of size.

However, we may encounter alternative definitions, such as $e^N$ or $N!$. For the purpose of our discussion and in line with common practice in computational complexity theory, we will deem these as \textit{unnatural} and exclude them from consideration, sidestepping any further explanations. Through this, we maintain a focus on the \textit{natural} definitions which are central to our understanding of computational complexity.

Given the definition of how to compute the hardness of a combinatorial problem by the timing an algorithm needs for solving it, we can ask as didactical example which of the three types of optimization problem, i.e., optimization, evaluation and decision, is the hardest.

In essence, when the cost of any configuration is determined in polynomial time, it implies that the evaluation problem's complexity does not exceed that of the optimization problem, i.e., optimization is harder than evaluation. This is because if we can identify the optimal configuration using polynomial time, deducing its cost is equally feasible within polynomial time. Moreover, when considering the decision problem (which is determining if there's a configuration with a cost less than $E_0$), its complexity is at most that of the evaluation problem. Therefore, when ranking them by increasing difficulty, it goes from decision to evaluation to optimization \cite{mezard2009information}.


\subsection{Polynomial Reduction and Complexity Classes}
\label{subsec::cpolynomial_reduction}
In the above section, we have discussed how making a ranking among the decision, evaluation and optimization problems. In this section, we focus on decision problems, and we would like to compare the levels of difficulty among them. For doing so, we introduce the notion of polynomial reduction, which formalizes the sentence "\text{not harder than}", which helps to have a classifications of decision problems.

To understand polynomial reduction, we must first recognize that it acts, in our case, as a bridge between decision problems, enabling us to use solutions from one problem to solve another. Simply put, if we have two decision problems, Problem A and Problem B, and we can transform any instance of Problem A into an instance of Problem B in polynomial time (by using any efficient algorithm, if it exists), we say that Problem A reduces to Problem B. This is polynomial reduction. This concept becomes particularly useful when Problem B is already solved or is well-understood. Generally speaking, if we can reduce Problem A to Problem B efficiently (in polynomial time), we can take any solution for Problem B and use it as a solution for Problem A \cite{goldreich2010p}. It is like translating a book into another language - if we can translate it quickly and someone else can read that language, then they can understand the book. Polynomial reduction also helps us in understanding the relative hardness of computational problems. If Problem A reduces to Problem B and we know that Problem B is hard, we can infer that Problem A is at least as hard as Problem B. So, polynomial reductions not only allow us to solve problems using solutions of other decision problems, but also enable us to establish a hierarchy of problems based on their hardness. This very hierarchy leads us to the concept of complexity classes, which categorize problems based on their computational complexity. They are our next port of call on this exploration of computational complexity.

Finally, we can introduce the concept of complexity classes. The classification of decision problems with respect to \textit{polynomiality} is as follows:
The \textbf{P Class}, also known as polynomial time, consists of decision problems that can be solved by an algorithm in polynomial time. Algorithms in the P class are considered efficient because they can solve problems with large inputs in a reasonable amount of time. Common examples of problems in the P class include sorting, searching, and matrix multiplication. The \textbf{NP Class}, which stands for nondeterministic polynomial time, contains problems for which a potential solution can be verified in polynomial time. In other words, if a solution is proposed, it can be checked or verified efficiently. However, finding the solution itself may not be computationally easy. For example, the decision version of the TSP is in NP: if there is a TSP tour with cost smaller than $E_0$, the verification is simple. Indeed, given the tour and its cost,  it can be computed in linear time, allowing one to check that it is smaller than $E_0$. The \textbf{NP-Complete class} encompasses some of the most challenging problems within the larger class NP. A decision problem is designated as NP-Complete if it meets two specific conditions: \textit{i}) The problem resides in the class NP. \textit{ii}) The problem is, in a sense, as \textit{hard} as the hardest problems in NP. This means that any other problem in NP can be reduced to it through a polynomial-time reduction \cite{goldreich2010p}. In essence, if you can solve this problem quickly (in polynomial time), you can solve any problem in NP quickly. The second condition essentially forms the crux of what makes a problem NP-Complete. A major achievement of the theory of computational complexity was obtain by Cook in 1971 \cite{cook2023complexity}. He showed that the satisfiability problem is NP-Complete. This allowed then to find a huge list of decision problems in such a class, by using Polynomial Reduction. 
We conclude the discussion on the complexity classes  introducing the  \textbf{NP-hard} class. This class contains those problems which, not being in NP, are at least as hard as NP-Complete problems. These include both decision problems that cannot be checked or verified efficiently, and non-decision problems. For examples, the optimization and the evaluation versions of the TSP are NP-hard \footnote{Our classification in complexity classes does not cover all the cases that today are present in the literature. We refer to \cite{goldreich2010p} the reader interested in.}.

One of the major open question in the theory of computational complexity is whether the classes P and NP are distinct or not. This perplexing puzzle has challenged researchers and academics for decades, as it lies at the heart of computational complexity theory. The resolution of this question could have profound implications for various fields, including cryptography, optimization, artificial intelligence, and algorithm design. The P=NP problem stands as one of the seven mathematical conundrums known as the Millennium Problems, each of which carries a million-dollar prize for its resolution \cite{carlson2006millennium}. If the P$=$NP conjecture were proven true, it would signify that any problem for which a solution can be checked efficiently could also be solved efficiently. This would have groundbreaking implications across various fields. For example, a positive resolution to P$=$NP would mean that many complex problems, currently believed to be computationally intractable, could be solved using efficient algorithms. This would revolutionize various industries, enabling faster and more accurate data processing, optimization, decision-making, and artificial intelligence. On the other hand, if P$\neq$NP is proven, it would confirm that there are problems in NP that cannot be efficiently solved, even if a proposed solution can be quickly verified. This has equally profound consequences. For example, the security of modern cryptographic systems relies on the assumption that certain problems are hard to solve, even for powerful computers. If P$\neq$NP, these cryptographic systems would remain secure, safeguarding sensitive data and communication. 
Almost the totality of computer scientists believe that P$\neq$NP. However, a small fraction of scientists still suspect that P$=$NP. In this small fraction, there is Donald E. Knuth. In a footnote on Volume 4, Fascicle 6, \textit{The art of computer programming: Satisfiability} \cite{knuth2015art}, he wrote: "\textit{At the present time very few people believe that P = NP [see SIGACT News 43, 2 (June 2012), 53–77]. In other words, almost everybody who has studied the subject thinks that satisfiability cannot be decided in polynomial time. The author of this book, however, suspects that $N^{O(1)}$-step algorithms do exist, yet that they’re unknowable. Almost all polynomial time algorithms are so complicated that they lie beyond human comprehension, and could never be programmed for an actual computer in the real world. Existence is different from embodiment}"\cite{knuth2015art}. We believe that this is a very useful encouragement to the scientific community. Knuth's perspective serves as a reminder that within the realm of complex and intricate problems like P=NP, the existence of alternative possibilities should not be dismissed hastily. It is an encouragement to keep open a door that seems to be closed.
Indeed, a century ago, the field of physics faced a similar situation. The scientific community was pondering the existence of the \textit{ether}, a hypothetical substance believed to permeate the universe and serve as the medium for the propagation of light. However, the Michelson-Morley experiment \cite{michelson1929repetition}, conducted in 1887, revealed null results, effectively falsifying the conjecture of the ether's existence. Similar to the era of ether, we find ourselves standing at a crossroads in computer science. The question of whether P=NP remains an enigma, waiting to be unraveled. As Knuth urges us to consider, conducting experiments and pushing the boundaries of our understanding is vital. Just as physicists sought empirical evidence to resolve the mystery of the ether, computer scientists must also embrace experimentation to shed light on the P=NP problem definitively. By fostering an environment of open-mindedness and a willingness to explore the unknown, we can strive to find the truth that lies hidden in the complex tapestry of computational complexity. While P$\neq$NP is the prevailing belief, we must not forget that breakthroughs have often emerged from unexpected places.

\section{Optimization and Statistical Physics}
\label{sec::optwithSM}

Up to this point, we have merely scratched the surface of computational complexity and the challenges posed by combinatorial optimization problems. However, the true depth and nuance of these topics begin to emerge when viewed through the lens of statistical mechanics. Statistical mechanics is a branch of physics that helps us understand how the tiny building blocks of matter, like atoms and molecules, behave in large groups \cite{huang2009introduction}. It seeks to bridge the microscopic interactions of individual particles with macroscopic observables, like temperature, pressure, and energy, through the use of statistical averages and probabilities. By treating particles as statistical entities, rather than considering each particle's individual motion, statistical mechanics provides a powerful framework for studying the thermal, mechanical, and dynamical properties of matter, as well as phase transitions and other collective phenomena \cite{baldovin2023ergodic}. The field plays a crucial role in explaining the thermodynamic behavior of materials and systems, shedding light on a wide range of phenomena in condensed matter physics, astrophysics, and many other disciplines, like computer science \cite{aurell2016diffusion,marino2023hard,marino2020large,marino2016advective,caracciolo2021criticality,caracciolo2018solution,caracciolo2017finite,caracciolo2019selberg,capelli2018exact,malatesta2019fluctuations,malatesta2017two,giorgini2020two,giorgini2022correlation,pittorino2017chaos,stucchi2021order,giambagli2022diffusion,chicchi2021reconstruction,marino2016entropy}. This perspective offers a vast landscape of fascinating insights and discoveries. 

Imagine you are in a world where optimization and statistical physics are intertwined in a harmonious dance. In this dance, an optimization challenge steps forward, its defining attributes being a specific set $\mathcal{X}$ of potential configurations and a cost function $E$ mapped out on this set with values reaching out into the real number plane, $\mathbb{R}$. In the waltz of optimization, the goal is clear: find the configuration $C \in \mathcal{X}$ that takes the smallest toll, the one with the least cost. But here is a twist in our dance. Instead of focusing solely on the smallest cost, we can also apply a Boltzmann probability measure to our configuration space. It is like casting a net of probabilities across our space, with each configuration $C$ assigned a specific probability based on a given $\beta$. In other words,  for a given $\beta$, every configuration $C$ is attributed a probability: $\mu_{\beta}(C)=\frac{1}{Z(\beta)} e^{-\beta E(C)}, \,\,\,\, Z(\beta)=\sum_{C \in \mathcal{X}}e^{-\beta E(C)}$ \cite{mezard1987spin,huang2009introduction,mezard2009information}. Here, the positive variable $\beta$ functions as an inverse temperature, and $Z(\beta)$ is the so called partition function.  When $\beta$ approaches infinity, the probability distribution $\mu_{\beta}$ concentrates on the configurations with minimum energy (colloquially referred to as ground states in the language of statistical physics). This limit is particularly relevant for optimization problems. However,  adopting the viewpoint of a statistical physicist, the landscape of our optimization problem broadens. We find ourselves investigating the traits of our distribution, $\mu_{\beta}$, even when $\beta$ remains finite. This is analogous to scrutinizing an object under variable lighting conditions, revealing different features as the light intensity changes. Monitoring the transformation of $\mu_{\beta}$ as we gradually increase $\beta$ often provides valuable insights, comparable to observing the myriad changes from sunrise to sunset. Examining the thermodynamic properties such as internal energy, entropy, among others, especially when these thermodynamic properties  exhibit gradual transitions, our methodology proves to be highly efficacious, rewarding us both from analytical and algorithmic perspectives. A notable tool at our disposal is the simulated annealing method \cite{kirkpatrick1983optimization,van1987simulated,schneider2007stochastic,mohseni2021nonequilibrium}, a formidable navigator of our configuration space. As we incrementally enhance $\beta$ values, akin to a seasoned explorer delving deeper into unknown terrains, this method persists in its quest until it uncovers the ground state.

We now turn our attention to this nexus of computation and physics, bridging the gap between these two seemingly disparate disciplines. The foundations of this connection are rooted in the shared mathematical structures and underlying principles that define both fields. Particularly, we will delve into the well-trodden realms of max-cut problems and spin glass, classic examples that demonstrate this compelling intersection. 

Max-cut problem \cite{commander2009maximum}, a combinatorial optimization problem defined in Sec. \ref{sec::COP}, shares an interesting connection with spin glass, a model system in statistical physics. The objective of the max-cut problem is to partition the nodes of a graph into two sets such that the sum of the weights of the edges between the sets is maximized. On the other hand, spin glasses \cite{mezard1987spin} are disordered magnetic systems with competing interactions, a perfect setting for analyzing optimization and disorder. These systems are described by the Hamiltonian $E(\mathbf{S}) = - \sum_{(ij)} J_{ij} S_i S_j$, where $J_{ij}$ are the interaction strengths (random variables), $S_i$ and $S_j$ are Ising spins $\{\pm 1\}$ \cite{ma1985statistical}. The spins are  located on the vertices of a graph and there are $N$ of  them. The sum $(ij)$ runs over all edges of the graph. Identifying the ground state of the corresponding spin glass, given the graph and the exchange couplings, embodies a typical optimization problem. Herein lies the intriguing parallel. The configuration that minimizes the energy of a spin glass system in statistical physics is akin to the optimal solution of a max-cut problem in combinatorial optimization. This analogy between the minimum energy state of a spin glass and the maximum cut of a graph forms a conceptual bridge between statistical physics and combinatorial optimization \cite{percus2006computational}. 

Our forthcoming discussion will explore these parallels in detail, shedding light on the striking ways in which statistical mechanics can elucidate complex computational quandaries. In doing so, we will not only probe the fascinating intricacies of max-cut problem and spin glasses but also underscore the profound interconnectedness of our scientific understanding. Imagine we have a collection of spins, which can take one of two states, splitting our set of vertices into two distinct subsets: $V_{\pm}=\{ i | S_i = \pm1 \}$ \cite{mezard2009information}. Let's visualize this as two sets, $V_{+}$ and $V_{-}$, each set hosting vertices with a specific spin state. Now, we will consider the set of edges that have one end-point in $V_{+}$ and the other in $V_{-}$. We will call this set of edges $ \gamma(V_{+})$. We can express the energy of a spin configuration using the equation $E(\mathbf{S})=\mathcal{C}+2\sum_{(i,j)\in \gamma(V_{+})} J_{ij}$,
where $\mathcal{C}=\sum_{(i,j)} J_{ij}$ is a constant summarizing all the exchange coupling values. When we talk about finding the ground state of the spin glass, it's like looking for the best way to divide our vertices into the two sets, $V_{+}$ and $V_{-}$, such that the sum of $c_{ij}=-J_{ij}$ over all edges $ \gamma(V_{+})$ is as large as possible. This is an instance of the max-cut problem. It is like trying to divide a graph into two node sets in such a way that the sum of weights of all respectively edges is maximized. In other words, we are seeking the division (cut) with the highest total maximal weight. Each cut corresponds to a set of edges, $ \gamma(V_{+})$, and is assigned a total weight $\sum_{(i,j)\in \gamma(V_{+})} c_{ij}$.

Once we have set up the max-cut problem, we can bring in some known conclusions from the literature. Usually, this problem is a tough nut to crack, falling under the NP-hard category, but there are exceptions. For certain types of graphs, we can actually solve it in polynomial time. A prime example is the case of a planar graph\footnote{A planar graph is a graph that can be drawn on a plane without any edges crossing. This means that you can flatten the graph onto a two-dimensional plane in such a way that its vertices and edges remain connected as in the original graph, but no two edges intersect each other except at their shared endpoint (a common node).} \cite{bollobas1998modern}. If we picture our vertices and edges on a plane, the max-cut can be calculated in polynomial time. This is equivalent to discovering an efficient route to the ground state of a spin glass arranged in a two-dimensional square lattice \cite{hartmann2011ground}. However, not all scenarios are as straightforward. Take a three-dimensional spin glass problem, for instance. This resides in the complex realm of NP-hard problems. Yet, even in this tough terrain, progress has been made \cite{baity2014janus}. In recent years, we have seen the development of efficient \textit{branch and bound} methods \footnote{The branch and bound method is an algorithmic approach used extensively in areas of decision making such as operations research, machine learning, and computer science, particularly in optimization problems. It is a systematic procedure for solving optimization problems by breaking them down into smaller sub-problems and using a bounding function to eliminate sub-problems that cannot contain the optimal solution. The method involves partitioning the problem into a series of smaller subproblems (branching) and then calculating an upper or lower bound of the optimal solution of each subproblem. If the calculated bound of a subproblem is worse than the current best known solution, that subproblem is discarded (bound), as it cannot contain the optimal solution. In this way, the branch and bound method reduces the search space, making the problem more computationally tractable. }\cite{caracciolo2023simulated}. These methods, like savvy mountain guides, help us navigate the problem based on its max-cut formulation. 

The connection between statistical mechanics and optimization is not only present between the max-cut and spin glass. Statistical mechanics has been  useful for understanding many other aspects of the complexity for  optimization problems. For showing such results we confine ourselves on only two combinatorial problem, the $k$-SAT and the Maximum Clique problem. We will define formally these two problem when it will be necessary, in such a way to help the reader in understanding easily the model and grasping immediately the connections between optimization and statistical mechanics. 

\subsection{The Satisfiability problem and phase transitions}\label{subsec::ksat}
The world of computational problems presents a wide range of challenges, among which the satisfiability problem, often referred to as SAT, holds a significant position. The SAT problem, in its most fundamental form, is defined by a set of $N$ Boolean variables. Each Boolean variable is binary, taking one of two values: true or false, typically represented by the values $1$ or $0$, respectively. Adding complexity to the SAT problem are $M$ constraints, each one forming a clause. A clause is a disjunction (an OR operation $\lor$) of literals, with each literal being either a Boolean variable or its negation. So, if we denote a Boolean variable as $x_i$, then both $x_i$ and $\neg x_i$ qualify as literals. Let's come back to our party scenario stated in the introduction. Imagine you're hosting a dinner party where guest preferences mirror our SAT problem's Boolean variables and clauses. You want to ensure everyone enjoys the evening, but some guests, for instance, prefer to sit next to certain others while avoiding some, a situation paralleling the Satisfiability problem. To bring this closer to home, consider a SAT problem with three guests: Alice ($x_1$), Bob ($x_2$), and Carol ($x_3$). Their seating preferences are: a)Alice ($x_1$) prefers sitting next to Bob ($x_2$) OR not near Carol ($\neg x_3$); b)Bob ($x_2$) does not mind sitting next to Alice ($x_1$) OR Carol ($x_3$); c)Carol ($x_3$) would like to sit next to Alice ($x_1$) OR not near Bob ($\neg x_2$). These preferences can be represented as a SAT problem in Conjunctive Normal Form (CNF):$F = (x_1 \lor x_2 \lor \neg x_3) \land (x_2 \lor x_1 \lor x_3) \land (x_3 \lor \neg x_2 \lor x_1)$,where the symbol $\land$ is the logical operator AND. An instance of a SAT problem is therefore a conjunction of clause. Our task is to find a satisfiable (SAT) assignment for $x_1$, $x_2$, and $x_3$ that makes all clauses true, that is, satisfies all guests. Given the dichotomous nature of Boolean variables, each clause could have $2^k$ potential assignments where $k$ is the number of literals in the clause. One possible SAT assignment might be $x_1 = 1$ (Alice is present), $x_2 = 1$ (Bob is present), and $x_3 = 0$ (Carol is absent). We can move Carol to another table. This SAT assignment satisfies all preferences, mirroring the aim of the SAT problem: to find an assignment that satisfies all clauses. 
However, not all assignments lead to satisfied clauses. An UNSAT assignment does not make the CNF formula true, indicating that not all guests' preferences can be accommodated. The broad SAT problem branches into specific subsets. In $k$-SAT, each clause is restricted to exactly $k$ literals, while the MAX-SAT problem aims to find an assignment that satisfies the maximum number of clauses. Different SAT problems belong to distinct complexity classes. The 2-SAT problem is part of class P, solvable in polynomial time, while 3-SAT is NP-Complete, indicating a higher complexity. The MAX-SAT problem falls into the NP-hard category. These problems are more than theoretical constructs; they find practical applications across multiple domains. In artificial intelligence, SAT is used for knowledge representation and automated reasoning \cite{bundy2013interaction}. In hardware and software verification \cite{clarke2004sat}, SAT solvers detect bugs and vulnerabilities. SAT also models molecular structures in bioinformatics \cite{altman1998bioinformatics}, plans and schedules tasks in operations research, solves puzzles, and even facilitates cryptographic attacks \cite{soos2009extending}.

Now, we embark on an exploration of the captivating world of the Boolean satisfiability problem using the powerful tools of statistical mechanics \cite{barthel2002hiding}. Our focus lies on examining random $k$-SAT problems, which offer fascinating insights into the behaviors of complex systems. The random $k$-SAT problem is viewed by statistical physicists as a problem exhibiting the so-called frustration property\footnote{In the context of spin glasses, frustration refers to the presence of competing interactions in the system which prevent it from reaching a clear or 'unfrustrated' ground state. This property is a key feature of spin glass systems and plays a crucial role in their complex behavior. More specifically, in a spin glass system, the interactions between spins (which can be thought of as tiny magnets) can be either ferromagnetic (favoring parallel alignment of spins) or antiferromagnetic (favoring antiparallel alignment of spins). When a spin is involved in both ferromagnetic and antiferromagnetic interactions with its neighbors, it is not possible to satisfy all these interactions simultaneously. This situation is referred to as \textit{frustration}.}, which is of great interest to physics of spin glasses, and, hence, it has enjoyed a great deal of attention in the statistical physics literature \cite{martin2001statistical,kirkpatrick1993statistical,biroli2002phase,monasson1997statistical,monasson1996entropy}. 
These intriguing problems involve $N$ variables and $M$ clauses, with each clause containing exactly $k$ distinct literals, and is picked up with uniform probability distribution from the set of $\binom{N}{k} 2^k$ possible clauses. Under these conditions, we delve into unraveling the probability of a problem being satisfiable, as we vary the clause density represented by the ratio $\alpha=M/N$.

\begin{figure}
    \centering
    \includegraphics[width=1\columnwidth]{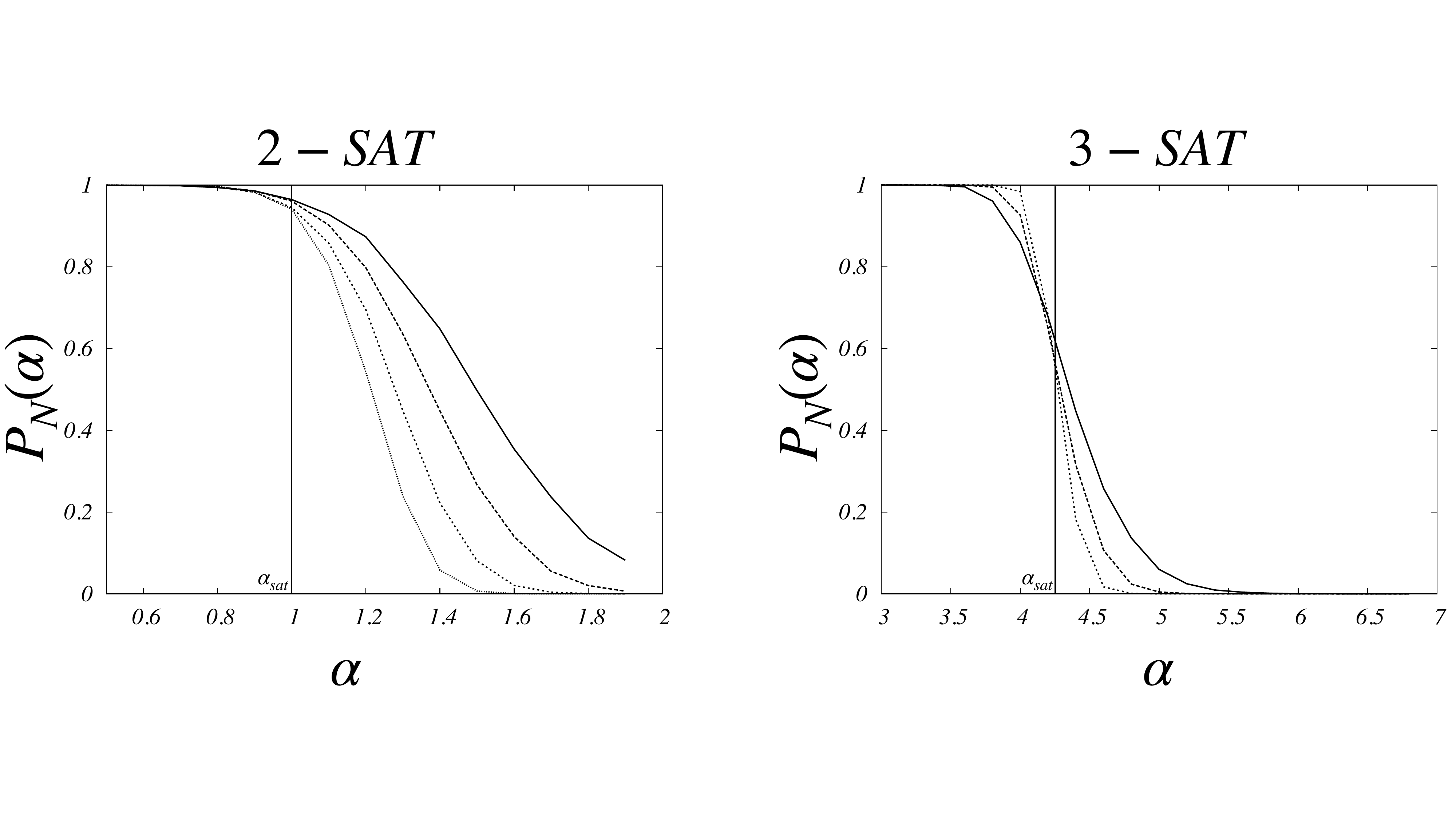}
    \caption{The probability that a randomly generated $k$-SAT formula is satisfied as a function of the parameter $\alpha$. The result for $2$-SAT is presented on the left and for $3$-SAT on the right. Each point is the result of averages on $10^4$ randomly generated formulas. The curves for $2$-SAT correspond to formulas of size $N=50, 100, 200$ and $400$ (from right to left). In the case of $3$-SAT, the curves correspond to $N=50, 100, 200$ (from right to left). The vertical lines identify the threshold value $SAT-UNSAT$, $\alpha_{sat}=1$ for $2$-SAT, and $\alpha_{sat}=4.267$ for $3$-SAT.}
    \label{fig:1}
\end{figure}

The exploration begins with a remarkable observation - at $\alpha=0$, every problem is satisfiable. This makes sense as we have fewer clauses to satisfy, leaving a wealth of possibilities to fulfill the conditions. However, as we increase $\alpha$ and approach infinity, something intriguing happens. The probability of a problem being satisfiable, $P(\alpha)$, diminishes until it ultimately vanishes in the limit $\lim_{\alpha \to \infty} P_N(\alpha)=0$. As we venture further into the realm of statistical mechanics, we encounter an enthralling phenomenon - phase transition \cite{10.5555/3070217.3070238}. With an expanding number of variables $N$, the behavior of $P_N(\alpha)$ takes on a striking transformation. The curves of $P_N(\alpha)$ become steeper and more pronounced, culminating in a step-function in the thermodynamic limit. This intriguing step-function signifies the partitioning of the space of solutions for the $k$-SAT problem into two distinctive regions. On one side $\alpha \in [0, \alpha_{sat}]$, we find problems with feasible solutions, while on the other $\alpha \in (\alpha_{sat}, \infty)$, no assignment exists to satisfy all clauses of a given instance. The presence of such phase transitions opens a captivating avenue of exploration into the intricate nature of problem-solving and the boundaries of computational complexity. Mathematical studies have confirmed the existence of these phase transitions for various values of $k$. For instance, the $2$-SAT problem showcases a critical SAT-UNSAT threshold value at $\alpha_{sat}=1$ \cite{mezard2009information}, while the recent discovery for the $3$-SAT problem places its threshold value at $\alpha_{sat}=4.267$ \cite{10.1145/2746539.2746619}. Remarkably, this behavior of the probability of finding a SAT assignment remains consistent across all different values of $k$ \cite{10.1145/2746539.2746619}.
To better grasp this enthralling topic, we present the Fig. \ref{fig:1}. This figure visually illustrates the trend of $P_N(\alpha)$ as a function of $\alpha$ for both random 2-SAT and 3-SAT problems. As the plot unfolds, we witness the fascinating transition from satisfiability to unsatisfiability, providing valuable insights into the underlying complexity of these problems. The exploration of phase transitions and the probability of finding satisfying assignments extends far beyond mere theoretical curiosity. These profound insights hold the potential to revolutionize algorithm design, computational efficiency, and our understanding of the universe's fundamental intricacies. As we navigate this captivating journey, we are reminded of the spirit of scientific inquiry that has guided great minds through the ages, inspiring us to push the boundaries of knowledge and embrace the beauty of the unknown.

Some of these great minds in \cite{monasson1999determining} reported an analytic solution and experimental investigation of the phase transition in random $k$-SAT. As stated above, depending on the given input parameters,i.e., the value of $k$, the computational complexity for this problem might follow an exponential or polynomial growth pattern in relation to the problem size. Interestingly, a stark, discontinuous transition is observed in the exponential growth scenario, while a continuous, or second-order transition, emerges in the polynomial case. This research elegantly establishes the nuanced relationship between phase transition phenomena and the complexity of typical computational cases. Continuous, or second-order phase transitions, result in a polynomial growth of resource requirements in line with the problem size. Conversely, discontinuous or random first-order phase transitions cause an exponential surge in resource requirements, emblematic of genuinely complex computational problems. These findings underscore the remarkable potential of statistical physics techniques in unearthing unique insights into computational phenomena, fostering the development of improved problem-solving methods. Moreover, this research introduces the novel $(2 + p)$-SAT model, a seamless bridge between the computationally manageable $2$-SAT and the elusive NP-Complete $3$-SAT. This model, by varying the value of $p$, provides fresh insights into the complexity of typical cases and straddles the boundary between polynomial and exponential problems. The $(2 + p)$-SAT model, when represented as a diluted spin glass model, delineates an order parameter that encapsulates the statistics of optimal assignments aiming to minimize clause violations, i.e., the magnetization. The distribution of magnetizations outlines the microscopic structure of the ground states, with an inflexible \textit{backbone} of highly constrained variables and a mid-distribution of less restricted variables. This rigid \textit{backbone} of frozen spins provides a logical explanation for the high cost associated with heuristic search near the phase boundary. The delineation of the backbone, coupled with the presence of critical fluctuations, exemplifies the \textit{two components} concept, analogous to the physics of glasses or granular materials. These findings not only deepen our understanding of computational complexity in the realm of $k$-SAT but also create fascinating intersections between computational models and physical system properties, thereby paving the way for wide-ranging interdisciplinary explorations.
The phase transition threshold holds a special fascination due to its association with the \textit{easy-hard-easy} patterns of computational cost \cite{zdeborova2016statistical,zdeborova2007phase}. It is relatively effortless to find a satisfying assignment at low $\alpha$ values, while it becomes substantially more challenging near the threshold, and interestingly, the difficulty eases off once again at higher $\alpha$ values. Even above the critical ratio $\alpha_{sat}$, demonstrating unsatisfiability remains a tough proposition, with a lower bound on the search cost that exhibits a power law decrease with $\alpha$. In this intricate landscape, researchers have found statistical mechanics to be a valuable ally in deciphering the enigmatic behavior of the solution space for the random $k$-SAT problem. In a notable study by \cite{krzakala2007gibbs}, the researchers uncovered the existence of distinct phase transitions within the solution space of a random constraint satisfaction problem. By examining the $k$-SAT problem and $q$-coloring problem of random regular graphs, they investigated the uniform measure grounded in the random set of solutions. In their exploration, they discerned that as the constraints per variable escalated, the uniform measure first disintegrated into an exponential number of pure states, or \textit{clusters}, before subsequently condensing over the largest such states. Strikingly, these transitions, decomposition, and clusterization are characterized by sharp transitions for typical large instances. Once these transitions were established, the researchers turned their attention to providing a formal definition of each phase transition. They did so in terms of different notions of correlation between distinct variables involved in the problem. The degree of correlation has a profound influence on the performance of many searching/sampling algorithms, enhancing the significance of their findings. For example, for the  random $k$-SAT problem, the space of solutions unravels intriguing transformations as the parameter $\alpha$ spans the entire SAT phase. At sufficiently low $\alpha$ values, the set of solutions are interconnected, forming a single cohesive cluster. As $\alpha$ escalates, it is not just that the number of solutions dwindles. At a critical $\alpha_d$, the random $k$-SAT ensemble experiences a transformative phase transition - the unified solution space shatters, fragmenting into an exponentially large number of clusters, each cluster bearing solutions that are distinct from others with a Hamming distance \cite{hamming1950error} of $O(N)$. Here, if an energy function $E$ is defined as the number of unsatisfied clauses in a configuration, an intriguing pattern emerges for $\alpha>\alpha_d$. The energy function encounters exponentially many local minima of positive energy, potentially ensnaring algorithms designed to find solutions through energy relaxation, such as Simulated Annealing. As $\alpha$ continues its upward climb, each cluster shrinks further, losing solutions along the way. Yet, the most dramatic change occurs in the number of clusters. Leveraging the cavity method \cite{mezard2003cavity} to count clusters based on their solution population, researchers have mapped out a detailed description, unearthing additional phase transitions. Notably, at a value $\alpha_c$, a condensation phase transition takes place, where for $\alpha>\alpha_c$, most solutions belong to a sub-exponential number of clusters. This transition creates formidable long-range correlations among variables in typical solutions, which are hard to approximate by any finite horizon algorithm. Generally, we observe that $\alpha_d<\alpha_c<\alpha_{sat}$. A significant portion of this portrayal of the solution space has been proven rigorously in the large $k$ limit \cite{achlioptas2008algorithmic,achlioptas2006solution}. Now, if we shift our focus to the algorithmic implications, a captivating question arises - does the rich structure of the solution space impact the performance of searching algorithms? Clustering at $\alpha_d$ might slightly impede algorithms that sample solutions uniformly, yet there exist many algorithms capable of finding at least one solution even for $\alpha>\alpha_d$.
It is a solid conjecture that the formula's complexity is tied to a subset of highly correlated variables, which pose a significant challenge when assigning them correctly. The most extreme scenario is a unique subset of variables. This concept, referred to as the \textit{backbone} earlier, when applied to solutions within a single cluster, led to the definition of \textit{frozen variables} \cite{molloy2018freezing}. These are variables that maintain the same value across all solutions within a cluster. It has been established that the fraction of frozen variables within a cluster is either zero or is lower bounded by $(\alpha e^2)^{-1/k-2}$. In cases where this holds, the cluster is termed as \textit{frozen}.
The conjecture suggests that finding a solution within a frozen cluster is a challenging feat, practically requiring an exponential time growth with $N$. Therefore, an efficient algorithm operating within polynomial time should ideally seek out unfrozen clusters for as long as they exist. But counting unfrozen clusters is no easy task, a challenge that is only been overcome quite recently for a simpler problem - bicolouring random regular hypergraphs \cite{braunstein2016large}. For random $k$-SAT, we only have partial results, defined in terms of two thresholds: for $\alpha>\alpha_r$ (rigidity), typical solutions are in frozen clusters (though a small fraction of solutions may still be unfrozen), while for $\alpha>\alpha_f$ (freezing), all solutions are frozen. It has been rigorously established that $\alpha_f<\alpha_{sat}$ strictly for $k>8$. For smaller $k$ values, which are more interesting for benchmarking solving algorithms, we know from the cavity method that $\alpha_r \sim 9.883(15)$ for $k = 4$, while for $k = 3$, the estimate $\alpha_f \sim 4.267(1)$ \cite{marino2016backtracking}. Generally, $\alpha_d<\alpha_r<\alpha_f<\alpha_{sat}$. The conjecture not only implies that no polynomial time algorithm can solve problems with $\alpha>\alpha_f$, but also suggests that even finding solutions close to the rigidity threshold $\alpha_r$ should be extremely difficult, given that unfrozen solutions become a tiny minority. This prediction aligns with the performance of all known algorithms.

\subsection{The Maximum Clique problem and the Overlap Gap Property}\label{subsec::clique}

The Maximum Clique Problem (MCP) \cite{bomze1999maximum} is a well-known combinatorial optimization problem that originates from graph theory. It finds numerous applications in diverse domains such as bioinformatics \cite{tomita2011efficient}, social network analysis \cite{pattillo2011clique}, computer science \cite{douik2020tutorial}, and operations research \cite{marino2023hard}. As we touched upon in the introduction using an illustrative example, imagine yourself in a social gathering, a party filled with an eclectic mix of people. As one among the $N$ guests, you have been tasked with a unique mission - to find the largest clique within the party, the biggest group of individuals who all know each other personally. Visualize drawing a map of the party, with each guest represented by a point and a line connecting those who know each other. This clique problem is akin to finding the most tightly-knit friend group in the party. While it may sound simple, even for a modest number of guests, the possibilities to explore multiply rapidly, and the problem becomes computationally challenging. This scenario captures the essence of the MCP in real-world terms. In a formal graph theory context, the term \textit{clique} refers to a subset of vertices in an undirected graph such that every pair of distinct vertices within the subset are connected by a unique edge. In other words, within a clique, every node is directly linked to every other node.

The MCP is defined as the computational task of identifying the largest possible clique within a given graph. Given a graph $\mathcal{G} = (\mathcal{V}, \mathcal{E})$, where $\mathcal{V}$ represents the set of vertices and $\mathcal{E}$ represents the set of edges, the goal of MCP is to find the maximum subset of vertices $K_n \subseteq \mathcal{V}$, such that for every pair of vertices $(v_1, v_2)$ in $K_n$, there exists an edge $(v_1, v_2)$ in $\mathcal{E}$. The size of a clique is measured by the number of vertices it encompasses, and a clique is considered \textit{maximum} if there is no other clique in the given graph that contains more vertices. It is important to note that the MCP is an NP-hard problem, indicating that there is no known algorithm that can solve all instances of the problem quickly (in polynomial time). Despite its computational complexity, the MCP continues to attract substantial research interest due to its theoretical significance and practical relevance to a wide array of real-world problems. 

The Erdős–Rényi graph is a renowned model in graph theory \cite{erdHos1960evolution}. Defined as $\mathcal{G}(N,1/2)$, this type of graph consists of $N$ nodes. An intriguing characteristic of Erdős–Rényi graphs is that each pair of nodes is connected with a probability of 1/2, with all pairs behaving independently. A fascinating question that often arises in the context of these graphs pertains to identifying the largest clique. In fact, the Erdős–Rényi graph is frequently employed as a prototypical model to compute the size of the largest clique.  The determination of $K_n$ in a random Erdős–Rényi graph provides a textbook illustration of the probabilistic method. This non-constructive mathematical proof technique involves establishing that a certain property holds merely by demonstrating that the probability of it holding is non-zero. Simple application of the probabilistic method can reveal with a high degree of certainty that $K_n$ approximates to $2 \log_2N$ for large $N$. However, the identification of such a maximum clique introduces a computational challenge that algorithmic complexity theory is yet to fully conquer. The field, established by pioneers like Richard Karp, hit a stumbling block in the form of an embarrassingly simple yet unbeatable algorithm. Karp proposed in his 1976 paper \cite{karp1976probabilistic} that a basic algorithm can locate a clique of about $\log_2N$ members, approximately half of the optimum size, and challenged the community to improve on this. The Karp algorithm for maximum clique commences by selecting an arbitrary site within the graph. Simultaneously, it discards approximately half of the sites that are not directly linked with the chosen site, thereby refining the pool of potential candidates. The algorithm then progresses to a remaining neighbor from the current frontier - the list of viable vertices yet to be evaluated for inclusion in the clique. Having picked a new vertex, it proceeds to eliminate approximately half of the remaining sites that do not connect with this newly chosen site. This iterative process of selection and discarding is maintained until the frontier has been completely exhausted, indicating that there remain no potential candidates to further extend the clique. Throughout the algorithm, the size of the frontier is halved at every step. This reduction indicates that the algorithm is unlikely to continue beyond $\log_2 N$ steps, where $N$ is the total number of nodes in the graph. Despite considerable advancements in algorithm design and implementation over the decades, the research community is yet to propose a substantially better alternative \cite{marino2023hard}. This inability to supersede an elementary algorithm reflects the first layer of the embarrassment associated with the MCP. Adding to the complexity is the absence of a robust theory of algorithmic hardness that can comprehensively explain why locating cliques of half-optimum size seems feasible, while finding larger cliques remains elusive for polynomial time algorithms. Traditional paradigms such as $P\neq NP$ and its variations offer little insight into this particular challenge. For example, assuming the widely believed conjecture that $P \neq NP$, finding a largest clique in a graph with $N$ nodes within a multiplicative factor $N^{1-\delta}$ is not possible by polynomial time algorithms for any constant $\delta \in (0,1)$ if $P \neq NP$ \cite{hastad1996clique}. This is a statement, however, about the algorithmic problem of finding large cliques in all graphs, and it is in sharp contrast with the factor $1/2$ achievable by polynomial time algorithms for random graphs $\mathcal{G}(N,1/2)$. Further complicating the narrative is the Hidden Clique Problem, a variant of the MCP introduced by Jerrum \cite{jerrum1992large}. This problem similarly exhibits a significant gap between the algorithmically achievable and the optimal. Such situations, where the optimal solution to an optimization problem involving randomness (denoted as $C$) significantly outperforms the best known polynomial time algorithm (denoted as $C_{ALG}$) leading to a noticeable $C_{ALG} < C$ gap, are surprisingly common. These gaps underline the fundamental challenges in bridging the difference between our theoretical understanding of problem complexity and our practical algorithmic capabilities.

In the quest to deeply comprehend the inherent computational complexity of optimization problems, Gamarnick masterminded a pivotal paper \cite{gamarnik2021overlap}. His work offers an innovative exploration of numerous theories rooted in the geometric landscape of the solution space of the core optimization problem. This revolutionary methodology was birthed in the field of statistical physics, more precisely, within the study of spin glasses \cite{mezard1987spin,krzakala2007gibbs}. Recently, this method has been amplified and applied to answer queries that extend far beyond the confines of statistical physics, touching upon challenges faced in the exploration of neural networks \cite{choromanska2015loss}. The cornerstone philosophy of this approach is the presumption that the intricate difficulties tied to the algorithmic hardness of a problem are reflected in the complex geometries of the optimal or near-optimal solutions \cite{marino2016backtracking}. The inaugural articulation of such an approach was geared towards decision problems, including random constraint satisfaction problems, a subject touched upon in the preceding section (see Sec. \ref{subsec::ksat}). Utilizing these valuable insights, Gamarnick \cite{gamarnik2021overlap} developed a comprehensive theory to decode the relationship between the hardness of a problem and the geometric configuration of the solution space tied to the base optimization problem. Although this theory bears resemblances to predictions built upon the clustering property, which suggests that the solution space of an optimization problem fragments into distinct clusters housing vastly different solutions, it introduces subtle yet vital nuances that allow for the exclusion of a vast array of algorithms, a feat previously deemed unattainable.

We shall dissect a generic optimization problem characterized by the expression $\min_{C} E(C, \Theta_N)$. Here, $E$ serves as the objective function awaiting minimization. On the other hand, $C$ signifies the solution, dwelling within the vast expanses of a high-dimensional space, $\Sigma_{N}$, where $N$ is the key that deciphers its dimensionality. This space is well-equipped with a metric (distance) $\rho(C,C')$, meticulously defined for each pair of solutions $C, C' \in \Sigma_{N}$. Accompanying the problem is its associated randomness, labeled as $\Theta_N$. Diving deeper into specifics, let's take the max-clique problem as an illustration. Here, $\Theta_N$ mirrors the random graph $\mathcal{G}(N, 1/2)$, and $C\in\{0,1\}^N$ serves as the encoding for a set of nodes knitting together a clique, where $C_i=1$ confirms the presence of node $i$ in the clique and $C_i=0$ negates it. Each unique instance of the problem, sprouted in accordance with the probability law $\Theta_N$, is represented by $\xi$. Here, the likelihood of birthing any specific graph stands at $2^{-\binom{N}{2}}$.
The objective function, $-E(C, \Theta_{N})$, keeps tally of the number of ones in $C$. As a contingency plan for the scenario where $C$ fails to encode a clique, we may set $E(C, \Theta_N)=\infty$ for such non-clique encoding vectors. As a result, for the largest clique problem, it can be anticipated that the random variable $C^*=-\min_{C\in \Sigma_N}E(C, \Theta_N)$ will hover around $2 \log_2 N$ with a considerable degree of likelihood.

Now, let's delve into the concept of the Overlap Gap Property (OGP), which is relevant to a specific instance $\xi$ of the randomness $\Theta_N$. We define OGP as follows: "\textit{The optimization problem $E_C(C, \xi)$ exhibits the OGP with values $\mu>0$, $0\leq \nu_1<\nu_2$, if for every two solutions $C, C’$ that are $\mu$-optimal in an additive sense (i.e., they satisfy $E(C, \xi)\leq C^* +\mu$ and $E(C’, \xi)\leq C^* +\mu$), it is the case that either $\rho_N(C,C’)\leq \nu_1$ or $\rho_N(C,C’)\geq \nu_2$}" \cite{gamarnik2021overlap}. To break it down in a simpler manner, this definition essentially means that any two solutions that are proximate (within an additive error of $\mu$) to the optimum solution are either close to each other (restricted within the distance $\nu_1$) or are relatively distant from each other (surpassing the distance $\nu_2$). This consequently uncovers a fundamental topological gap in the set of distances between solutions that lie near the optimal value.

In the case of random instances denoted by $\Theta_N$, the problem is said to manifest the OGP if the problem $\min E(C, \xi)$ exhibits the OGP with a high probability when $\xi$ is generated in accordance with the law $\Theta_N$. The concept of \textit{overlap} comes into play here, implying that distances in a normalized space can be directly associated with inner products, a terminology often referred to as overlaps in the realm of spin-glass theory. This is illustrated through the equation $|| C-C'||^2=||C||^2+||C'||^2-2<C,C’>$, where $<\cdot>$ symbolizes the inner product. In real-world scenarios, the solutions $C$ and $C'$ frequently exhibit identical or nearly identical norms, further reinforcing the overlap concept. The Overlap Gap Property (OGP) commands significant interest for particular combinations of parameters $\mu, \nu_1, \nu_2$, these parameters, however, are notably problem-dependent. This avant-garde approach, which underscores the topological disjointedness of overlaps among solutions that are near-optimal, has both theoretical and pragmatic implications. It presents a firm methodology to meticulously establish restrictions for specific categories of algorithms. These limitations originate from the affirmation of algorithmic stability, effectively suggesting that these categories of algorithms are intrinsically unable to surmount the gaps present in the overlap structures. Despite the general approach to eliminate these categories of algorithms being fairly problem-agnostic, the exact nature of the stability analysis required and the OGP itself are highly problem-dependent.

\section{Heuristic Algorithms and Learning Algorithms}

The concluding section of this chapter is dedicated to the exploration of algorithms. We will enumerate the primary strategies leveraged in solving combinatorial optimization problems, followed by a discussion on the invaluable insights that statistical mechanics has furnished for the creation of novel algorithms. In addition, we will venture into the domain of artificial intelligence (AI), presenting achievements garnered in AI, which currently appear to herald a promising new frontier for the resolution of hard problems.

\subsection{Heuristic algorithms: from greedy to message passing algorithms}

In the realm of computational complexity, heuristic algorithms \cite{cormen2022introduction,schneider2007stochastic} are powerful techniques used to tackle challenging computational problems, especially those within the NP class and NP-Complete problems. These approaches provide strategies for finding approximate solutions or exploring the solution space efficiently. Heuristic algorithms are problem-solving techniques that aim to find good solutions, even if they are not guaranteed to be optimal. These algorithms employ various rules, strategies, or domain-specific knowledge to guide the search for solutions. Unlike exact algorithms that guarantee the best solution, heuristics sacrifice optimality for efficiency, allowing for the exploration of large problem instances in a reasonable amount of time. Heuristic algorithms are particularly useful for combinatorial optimization problems. They often rely on iterative improvement, exploring neighboring solutions and gradually refining them towards better solutions. Examples of heuristic algorithms include greedy algorithms and local search algorithms (e.g., Simulated Annealing, Parallel tempering \cite{earl2005parallel}), evolutionary algorithms (e.g., Genetic Algorithms \cite{forrest1996genetic}, Ant Colony Optimization \cite{dorigo2006ant}), swarm intelligence algorithms (e.g., Particle Swarm Optimization \cite{kennedy1995particle}), message passing algorithms (e.g., Belief Propagation \cite{yedidia2003understanding}, Survey Propagation \cite{mezard2002analytic,braunstein2005survey}) and much more \cite{cormen2022introduction}. These algorithms play a crucial role in solving real-world instances of challenging problems, where finding exact optimal solutions is computationally infeasible. They provide near-optimal solutions and are widely used in various domains, including scheduling, routing, resource allocation, and machine learning. The literature on heuristic algorithms is so huge that it is also difficult to give some simple examples. For this reason we decided to focus on two heuristics: greedy and message passing algorithms. These algorithms are useful examples to understand how a combinatorial optimization problem can be attacked. For a more in-depth and specialized study of this topic, readers are encouraged to refer to \cite{kochenderfer2019algorithms}.

A greedy algorithm is a solution approach to optimization problems that hinges on a sequence of choices. At each decision point, the algorithm opts for the solution that seems best at that moment. However, it should be noted that this heuristic strategy does not always guarantee an optimal solution. The design of greedy algorithms generally follows a sequence of steps: 1) Express the optimization problem in such a way that, once a choice is made, there remains only one sub-problem to solve. 2) Prove that there always exists an optimal solution to the original problem that follows the greedy choice, so the greedy choice is always safe. 3) Verify the property of optimal substructure, that is, show that, after making the greedy choice, what remains is a subproblem with the property that combining an optimal solution to the subproblem with the greedy choice made yields an optimal solution to the original problem.

Determining whether a greedy algorithm will solve a particular instance of an optimization problem has no general method, but the properties of the greedy choice and the optimal substructure are the two key ingredients. If we demonstrate that a problem satisfies these properties, then we are on the right track to develop a greedy algorithm for that specific problem. At the heart of greedy algorithms is the principle of making locally optimal choices at each step in the hope of achieving a global optimum. This means we make decisions based on the current scenario without considering the implications of future subproblems. Therefore, while tackling an issue using a greedy approach, the algorithm selects the best seeming option at that given moment and proceeds to resolve the problems emerging from that particular decision. It is important to note that a greedy algorithm's decision could hinge on previous choices, however, it never depends on any future selections. As an example, we present the Karp algorithm \cite{karp1976probabilistic}. The Karp algorithm for maximum clique is an excellent embodiment of a greedy algorithm's primary strategy, as it optimizes decision-making at each step based on the current context. The algorithm's approach to solving the optimization problem is quintessentially greedy; it involves a sequence of choices, each following the same two-step process: selection and elimination.
The process commences by choosing a random site in the graph, reflecting the first step of expressing the problem in such a way that once a choice is made, there remains only one sub-problem to solve. This selection effectively sets the stage for the subsequent optimization process.
Following the greedy strategy, the algorithm proceeds to eliminate roughly half of the sites that do not have a direct link to the selected site. This step reflects the notion of a \textit{greedy choice}, where a locally optimal decision is made by refining the pool of potential candidates based on the current scenario. This choice is safe in the sense that it ensures an optimal solution can be derived from this stage.
As the algorithm progresses, it moves onto another neighbor from the current frontier, marking another greedy choice. Akin to the previous round, approximately half of the remaining sites not connecting to the newly selected site are discarded. Here, the algorithm follows the property of optimal substructure, leveraging the output from the prior greedy choice to solve the next subproblem.
This iterative process of selection and elimination - the algorithm's very core - is carried out until the frontier is entirely exhausted, implying no further candidates can extend the clique. Notably, the algorithm's choices are rooted solely in the current problem context, without considering solutions of future subproblems - a principle fundamental to greedy algorithms. Hence, the Karp algorithm's strategy beautifully showcases the basic principles of a greedy approach to solve optimization problems.

Message-passing algorithms \cite{sunil2012message} serve as a prominent pillar in the realm of graphical models, particularly in many random constraint satisfaction problems. Generally regarded as heuristic algorithms, these computational methods are primarily designed to offer approximate yet practically sufficient solutions \cite{mezard2009information}. The fundamental essence of heuristic algorithms is their proficiency in navigating vast solution spaces and attaining acceptable results within a reasonable timeframe, a characteristic that message-passing algorithms duly embody. However, in contrast to common heuristics, message-passing algorithms possess an additional intriguing property: they transition from being heuristic to exact when applied to tree-like structures. This implies that on tree-like graphical models, message-passing algorithms can derive exact inference solutions, making them a remarkably versatile tool in the landscape of computational algorithms. Despite their heuristic nature on graphs with loops, where convergence is not always guaranteed, message-passing algorithms continue to stand as a reliable method for a diverse array of problem-solving scenarios, paving the way for the development of advanced variations to handle more complex graphical structures. Message-passing algorithms operate on \textit{messages} associated with the edges of the factor graph \cite{loeliger2004introduction} underlying the inference-optimization problem and update them recursively through local computations performed at the vertices of the graph. A factor graph is a bipartite graphical representation widely used in  computer science and physics to describe the factorization of functions, notably within the context of probabilistic models. Essentially, a factor graph encapsulates a decomposition of a global function into a product of local functions or factors. It is composed of two types of nodes: variable nodes and factor nodes. Variable nodes represent the variables involved in the function, while factor nodes correspond to the local functions or factors. Edges in a factor graph connect factor nodes to their respective variable nodes, visually encapsulating how the global function breaks down into its constituent factors. 
\begin{figure}
    \centering
    \includegraphics[width=1\columnwidth]{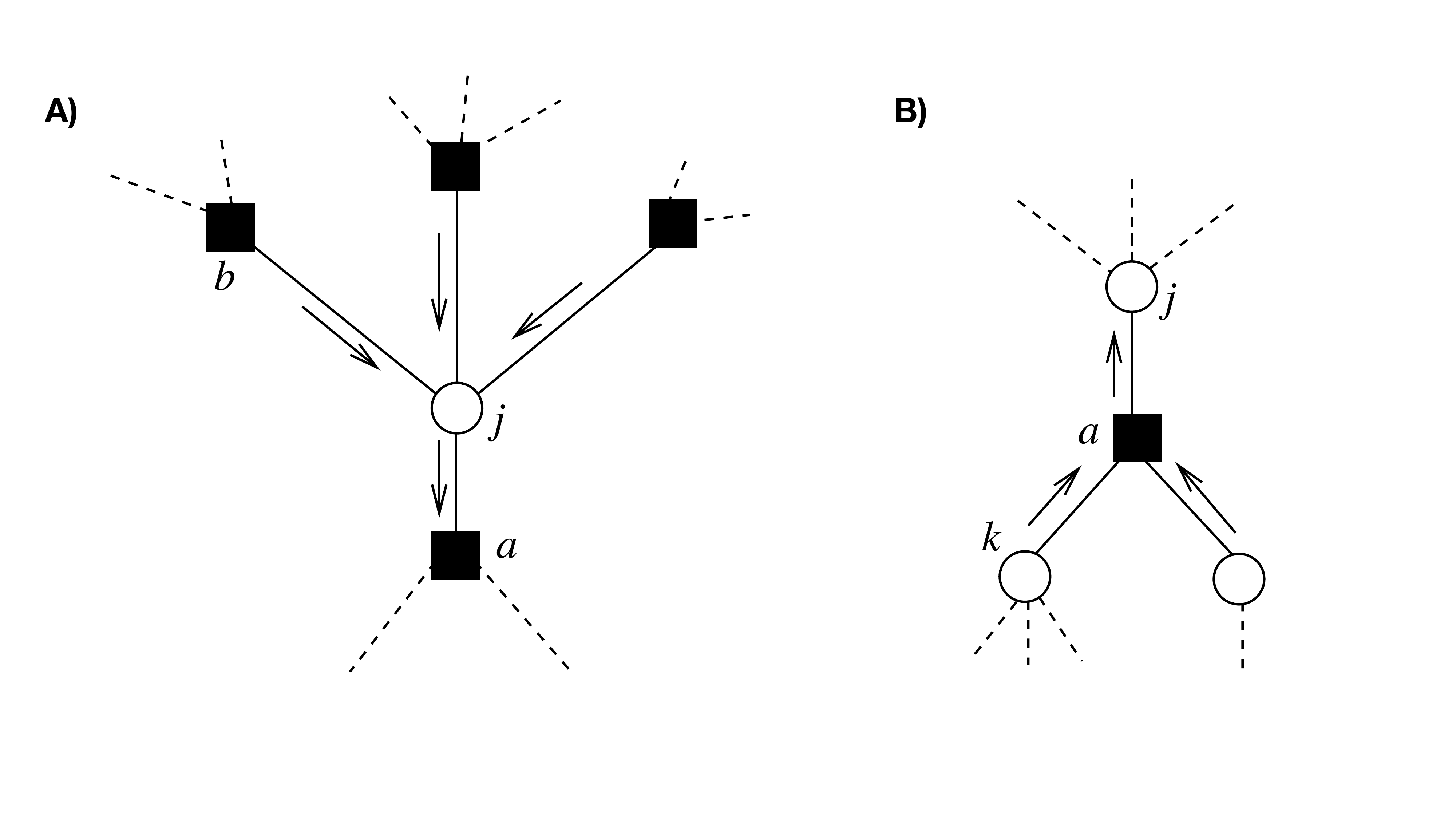}
    \caption{\textbf{A}: cartoon of a section  of a factor graph that participates in calculating $m^{(t+1)}_{j\to a}$. This message is influenced by the incoming messages that come from the factor nodes, except the one that comes from $a$. \textbf{B}: the section of the factor graph that is engaged in the calculation of $m^{(t)}_{a \to j}$. This message is influenced by the incoming messages that come from the vector nodes, except the one that come from $j$.}
    \label{fig:2}
\end{figure}
Message-passing algorithms stand out as a class of iterative procedures, the focal point of which are the \textit{messages} tied to directed edges on a factor graph. Each edge, denoted as $(i,a)$, where $i$ symbolizes a variable node and $a$ represents a function node, is associated with a pair of messages, $m_{i \to a}^{(t)}$ and $\hat{m}_{a \to i}^{(t)}$, at the $t$-th iteration stage. These messages are distinctive in that they carry values in the realm of probability distributions over their respective supports. Interestingly, when applied to a tree structure, the messages gradually converge as $t$ approaches infinity, settling at a fixed-point value. This convergence enables the calculation of marginal distributions on each variable node within an altered graph that excludes the factor $a$.  The transformation of messages is conducted through local computations at the nodes of the factor graph. When we speak of \textit{local}, we refer to how a specific node updates the outgoing messages based on the ingoing ones from the prior iteration (see to Fig. \ref{fig:2}). This is a characteristic trait of message-passing algorithms, and the variations amongst different algorithms within this family often lie in the exact formulation of their update equations.  At its core, a message-passing algorithm, as described in \cite{mezard2009information}, operates on a given factor graph through a sequence of systematic stages: 1) Firstly, an alphabet of messages, denoted as M, is established. This can be either continuous or discrete. The algorithm then begins to work with messages $m_{i \to a}^{(t)}$, $\hat{m}_{a \to i}^{(t)}$ which belong to the set $M$. These messages are tied to the edges within the factor graph. 3) Secondly, we define update functions, namely $\Psi_{i \to a}: M^{|\partial i\setminus a|} \to M$ and $\Phi_{a \to i}: M^{|\partial a \setminus i|} \to M$. These functions delineate the procedure for message updates. In this context, the symbol $\partial a$ incorporates all variables involved in constraint $a$, while $\partial i$ includes all factors associated with variable $i$. The $\setminus$ symbol denotes the difference with the respective element not implicated in the update rule. 3) Next, an initialization of the messages is performed. This could be a random initialization, represented as $m^{(0)}_{i\to a}$ and $m^{(0)}_{a \to i}$. 4) The final stage involves a decision rule. This refers to a local function that maps messages to a decision space, from which we aim to make a selection. Since our primary interest lies in computing marginals, we often consider the decision rule as a family of functions represented as $\hat{\Psi}_i:M^{|\partial i|} \to \mathcal{M}$, where $\mathcal{M}$ denotes the set of measures over $\mathcal{X}$, the alphabet of variables.

It is crucial to note a characteristic trait of message-passing algorithms: the messages that depart from a node are functions of the messages arriving at the same node via other edges. This property underscores the fundamental operating principle of these algorithms. We remaind to specilized book, for examples \cite{mezard2009information}.  As stated before, examples of message-passing algorithms are the Belief Propagation and the Survey Propagation. Belief Propagation (BP) is an algorithm designed to extract marginal probabilities for each variable node on a factor graph. Though it is exact on tree structures, it has also been found to be effective on loopy graphs. BP operates as an iterative message-passing algorithm, facilitating the exchange of messages from links to nodes. It subsequently computes marginal probabilities for each variable node. When the marginal probability is established, and BP has converged, a solution to the problem can be obtained by sampling from these predicted marginal probabilities. However, if the graph does not resemble a tree locally, BP might converge to a random and uninformative fixed point, marking a failure in the algorithm's operation \cite{yedidia2003understanding}. Despite this, BP has been successful in solving diverse combinatorial optimization problems. Interestingly, its application to the random $k$-SAT problem falls short far from the SAT-UNSAT transition \cite{mezard2009information}.
To address this issue, Mezard, Parisi, and Zecchina \cite{mezard2002analytic} introduced the Survey Propagation (SP) algorithm. It was developed under the assumption of one-step replica symmetry breaking and the cavity method of spin glasses. The algorithm proves particularly effective for low $k$ values and outperforms many existing methods in terms of running times and the sizes of instances it can handle \cite{marino2016backtracking}. SP operates on the factor graph with an underlying Conjunctive Normal Form (CNF) formula \cite{mezard2009information}. For larger $N$, SP is conjectured to perform better, as it navigates locally tree-like factor graphs, and cycles within the graph are $O(\log N)$. A detailed exposition of the SP algorithm is available in \cite{mezard2002analytic,braunstein2005survey}.
Generally, SP facilitates the exchange of messages between the $N$ variables and $M$ clauses to estimate the value that each variable should be set to in order to satisfy all clauses. More specifically, a message from SP, known as a survey, passed from one function node $a$ to a variable node $i$ (connected by an edge) is a real number $m_{a\to i} \in [0, 1]$. Under the assumption that SP operates over a tree-like factor graph, these messages adopt a probabilistic interpretation. Particularly, the message $m_{a\to i}$ corresponds to the probability that the clause $a$ sends a warning to variable $i$, suggesting the value that the variable $i$ should assume to satisfy itself \cite{parisi2003probabilistic,maneva2007new}.

\subsection{Learning algorithms: the GNN}

Artificial Intelligence (AI) has revolutionized numerous domains of human endeavor, with its capacity to handle complex tasks and vast amounts of data. A significant subset of AI is machine learning, which involves algorithms that can learn from and make decisions or predictions based on data. Deep Learning (DL), a further subset of machine learning, has emerged as a powerful tool due to its ability to construct and train neural networks with many layers, hence the term \text{deep}. Deep Learning has been exceptionally successful in numerous areas, particularly those involving large and complex datasets \cite{marino2023solving,marino2021learning,baldassi2021unveiling,baldassi2022learning,lucibello2022deep,giambagli2021machine,buffoni2022spectral,chicchi2021training,chicchi2023recurrent,marino2023phase}. One of its most recent achievements is the ground-breaking success of DeepMind's AlphaFold in "\textit{solving}" the protein folding problem, a long-standing challenge in the field of biology. DeepMind has used DL to predict the 3D structure of a protein based solely on its amino acid sequence \cite{jumper2021highly}. This breakthrough serves as a testament to the transformative power of AI in addressing computationally intensive and complex problems.
Following on from this, one exciting and emerging area of DL is the application of Graph Neural Networks (GNNs) \cite{wu2020comprehensive} to tackle combinatorial optimization problems. Graph Neural Networks are a special kind of neural network designed to work directly with data structured as a graph. GNNs are used in predicting nodes, edges, and graph-based tasks. Each layer of the GNN is a function $f[\cdot]$ with parameters $\mathbf{W}$ that takes the node embeddings and adjacency matrix and outputs new node embeddings.  This unique architecture makes GNNs particularly adept at handling combinatorial problems, where the solution's quality often depends on the entire configuration rather than individual components.
In the context of combinatorial optimization, GNNs can learn to predict the qualities of different configurations or even directly predict the solutions. They are trained on a set of problem instances with known optimal solutions, and through the learning process, they identify patterns and structures that can be generalized to unseen instances. The burgeoning debate in the field of artificial intelligence has recently raised the question of whether Graph Neural Networks (GNNs) can rise to the challenge and supersede the performance of state-of-the-art classical algorithms such as greedy or message-passing strategies. This discourse centers around the implications of the Overlap Gap Property (OGP), as discussed in Section \ref{subsec::clique}, a feature exhibited by a multitude of combinatorial optimization problems. Prominently, Gamarnik \cite{gamarnik2023barriers} has asserted that the presence of OGP significantly narrows the possibility for GNNs to surpass existing algorithms. In fact, he elucidates a fundamental constraint for deploying GNNs on random graphs, a constraint that holds true across a diverse spectrum of GNN architectures. Remarkably, this limitation pertains when the GNN's depth remains constant and does not scale proportionally with the size of the graph. This holds true irrespective of other facets of the GNN architecture such as the internal dimension or update functions. It is worth mentioning, however, that this does not preclude GNNs from being a valuable tool in the AI arsenal. While these constraints might limit the potential of GNNs in certain scenarios, the flexibility and learning capability of GNNs make them a promising approach for tasks where classical algorithms might struggle, particularly when the data or the problem setup is dynamic or when the structure in the graph data is complex and non-trivial. Thus, the choice between classical algorithms and GNNs might be more about leveraging the strengths of each approach for different types of problems rather than seeking a one-size-fits-all solution. It underscores the beauty of the diverse array of techniques we have in our hands to tackle the intricate landscape of combinatorial optimization problems.

\section{Conclusion}

This chapter has undertaken an in-depth exploration of the profound complexities inherent in solving combinatorial optimization problems. The journey began with the foundational definitions and concepts of graph theory, a crucial underpinning for the understanding of these challenges. We then embarked on an exploration of the intricate complexities encapsulated in the combinatorial optimization landscape. The presence of these complexities underscores the arduousness involved in the navigation and resolution of these problems. We then shifted our lens towards the intriguing nexus between statistical mechanics and combinatorial optimization problems. This symbiosis exemplifies how the methodologies from one field can help unlock solutions in another, as statistical mechanics has demonstrated with its robust toolbox for addressing combinatorial optimization problems. Central to our exploration was the elucidation of the Overlap Gap Property (OGP) of Gamarnik, a formidable barrier encountered by many algorithms when grappling with random constraint satisfaction problems. The ubiquity of OGP in a plethora of these problems sheds light on its pivotal role in impeding progress in resolving these issues. This insight facilitates a deeper understanding of the structural obstacles that need to be overcome to effectively handle combinatorial optimization problems. Moreover, we delved into the universe of heuristic algorithms, particularly greedy and message-passing algorithms. These methods stand as testament to the intriguing balance between simplicity and power, offering practicable solutions in reasonable timeframes despite their inherent approximation. Towards the end, we pivoted towards the promising frontier of machine learning algorithms, emphasizing on Graph Neural Networks (GNNs). These models embody the spirit of modern computation,i.e., adaptive, versatile, and capable of learning from data. Despite the inherent limitations highlighted by the presence of OGP, the inherent power of GNNs to capture complex patterns and dependencies in graph-structured data suggests that they will continue to be an area of active research and application. Overall, the multifaceted nature of the combinatorial optimization problems and the diverse array of approaches that we have explored to tackle these problems underline the richness and the challenges of this field. As we continue to push the boundaries of our understanding and our computational capabilities, the journey towards solving these complex problems remains an exciting and open journey of discovery.

\section*{Acknowledgments}

R.M. thanks Scott Kirkpatrick for the first reading of this chapter and for useful discussions.
This work is supported by \#NEXTGENERATIONEU (NGEU) and funded by the Ministry of University and Research (MUR), National Recovery and Resilience Plan (NRRP), project MNESYS (PE0000006) "A Multiscale integrated approach to the study of the nervous system in health and disease" (DN. 1553 11.10.2022).

\bibliographystyle{unsrt}
\bibliography{biblio}

\end{document}